\documentclass{aa}
\usepackage{txfonts}
\usepackage{graphicx}
\begin{document}
\title{Caught in the act: a helium-reionizing quasar near the line of
sight to Q0302$-$003\thanks{Based on observations collected at the
European Southern Observatory, Chile, (Proposals 66.A-0307 and
68.A-0194)}}

 \author{P. Jakobsen\inst{1}
 \and
R. A. Jansen\inst{1,2}
\and
S. Wagner \inst{3}
 \and
D. Reimers \inst{4} }

\offprints{P. Jakobsen, \hfill\break \email{pjakobsen@rssd.esa.int}}

 \institute{
Astrophysics Division, RSSD, European Space Agency, ESTEC, NL-2200 AG
Noordwijk, The Netherlands
  \and
Department of Physics and Astronomy, Arizona State University, Tempe
AZ 85287-1504, USA
\and
Landessternwarte Heidelberg, K{\"o}nigstuhl 12, D-69117 Heidelberg,
Germany
  \and
Hamburger Sternwarte, Gojenbergsweg 112, D-21029 Hamburg, Germany}

\date{Received 23 October 2002 / Accepted 30 October 2002}

\abstract{We report the discovery of a quasar at $z=3.050\pm0.003$,
closely coincident in redshift with the isolated low-opacity feature
seen near $z\simeq 3.056$ in the otherwise black portion of the
\ion{He}{ii} Gunn-Peterson absorption trough seen toward the $z=3.286$
background quasar Q0302$-$003, located $6\farcm5$ away on the sky. We
explore plausible models for the \ion{He}{iii} ionization zone created
by this neighboring quasar and its interception with the line of sight
toward Q0302$-$003.  At its present brightness of $V\simeq 20.5$ and
separation of $D_\perp\simeq3.2$~Mpc, the quasar can readily account
for the opacity gap in the \ion{He}{ii} absorption spectrum of
Q0302$-$003, provided it has been active for $t_Q\ga 10^7$~y. This is
the first clear example of the `transverse'  proximity effect and the
association of a quasar with its imprint on the intervening absorption
detected along an adjacent line of sight.
 
}

\titlerunning{A helium-reionizing quasar near Q0302$-$003}
\maketitle


\section{Introduction}

One of the more unique contributions of UV space astronomy to
observational cosmology has been the first detections of redshifted
intervening singly ionized \ion{He}{ii} Gunn-Peterson absorption due
to the Lyman forest and diffuse intergalactic medium (IGM) in the
far-UV spectra of high redshift quasars obtained with HST (Jakobsen et
al.\ 1994; Tytler et al.\ 1995; Reimers et al.\ 1997), HUT (Davidsen
et al.\ 1996) and FUSE (Kriss et al.\ 2001). These detections of
singly ionized intergalactic helium have not only confirmed
qualitatively a fundamental tenet of Big Bang nucleosynthesis theory,
but also provided direct support for the long-standing picture of a
photoionized intergalactic medium that was reionized by quasars and
young galaxies.

Although \ion{He}{ii} absorption has to date still only been observed
toward a total of four high redshift quasars, the dramatic change in
character and strength of the \ion{He}{ii} opacity with redshift
strongly suggests that the final stage of reionization of the universe
from \ion{He}{ii} to \ion{He}{iii} may have occurred at $z\simeq2.9$;
i.e.\ considerably later than hydrogen reionization which recent
evidence suggests may have taken place at $z\simeq 6-7$ (Becker et al.
2001; Djorgovski et al. 2001; Pentericci et al. 2002). Taken together,
these results imply that the hydrogen content of the IGM was reionized
early by sources having a relatively soft ionizing spectrum (i.e.\ the
first generations of stars), whereas helium reionization was only
completed later once the more energetic quasars and AGNs had formed at
$z\la 3$.

The \ion{He}{ii} absorption at redshifts $z\ga 2.8$ in the three lines
of sight  probed so far at the resolution and sensitivity of HST/STIS
(Q0302$-$003 ($z=3.29$), PKS1935$-$692 ($z=3.18$) and HE2347$-$4342
($z=2.89$)) all display an underlying trough of seemingly black
($\tau_\mathrm{He\,II} > 3$) absorption interrupted by one or more
isolated megaparsec-sized `voids' or `gaps' of finite
($\tau_\mathrm{He\,II} \sim 1$) opacity material (Reimers et al. 1997;
Anderson et al. 1999; Heap et al. 2000; Smette et al. 2002). This is
in marked contrast to the \ion{He}{ii} absorption at redshifts $z \la
2.8$ observed with HUT toward HS1700$+$642 ($z=2.72$) and FUSE toward
HE2347$-$4342, which  is much weaker and matches that of the
\ion{H}{i} Lyman forest more closely (Davidsen et al. 1996; Kriss et
al. 2001).


\begin{figure*}
\centerline{
\resizebox{0.85\hsize}{!}{\includegraphics{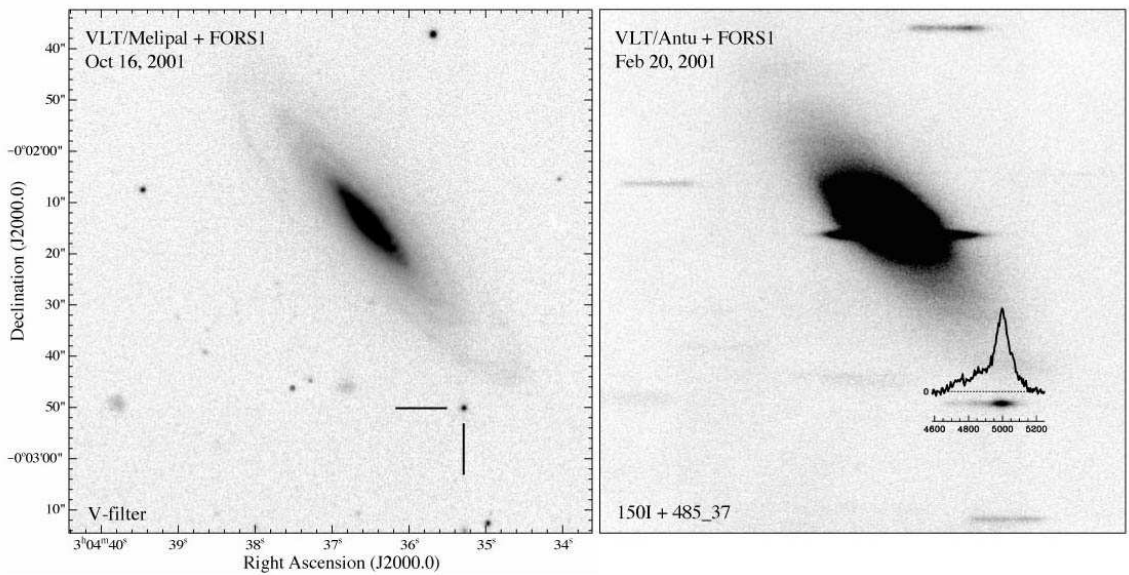}}
}
\caption{ $1\farcm7 \times 1\farcm7$ portion of a 120~s VLT/FORS
$V$-filter image (left) and the corresponding segment of the 1665~s
filtered slitless spectrographic exposure (right) showing the location
and discovery spectrum of QSO\allowbreak\,03020$-$0014, located
$6\farcm5$ northwest of Q0302$-$003. The 150I grism crossed with
the 485\_37 filter was used to isolate Ly$\alpha$ in the redshift range
$2.8<z<3.2$. The bright $B\simeq15.5$ Sb galaxy in the foreground is
MCG$+$00$-$08$-$082}
\label{discovery}
\end{figure*}


As discussed in detail by Reimers et al.\ (1997), Heap et al.\ (2000)
and Smette et al.\ (2002), the isolated \ion{He}{ii} opacity gaps seen
at $z \ga 2.8$ correspond closely to clearings seen in the
 \ion{H}{i} Lyman forest, and are naturally interpreted in
the conventional quasar reionization picture as the not yet
overlapping \ion{He}{iii} ionized regions surrounding other foreground
quasars or AGNs located near the line of sight.

The four \ion{He}{ii} voids seen toward Q0302$-$003, HE2347$-$4342,
and PKS1935$-$692 all appear at redshifts between $z\simeq 2.81$ and
$z\simeq 3.10$, and display widths in the range $\Delta z\simeq 0.01 -
0.02$. In the currently fashionable $H_0=65$~km~s$^{-1}$~Mpc$^{-1}$,
$\Omega_0=0.3$, $\Lambda_0=0.7$ cosmology, these voids extend
distances of $D_\parallel\simeq 2-5$~Mpc along the line of sight,
corresponding to transverse angular distances on the sky in the range
$\theta\simeq 5^{\prime}-10^{\prime}$. That this angle is comparable
to the field of view of available multi-object spectrographs, makes it
obvious to attempt to search for these putative helium-ionizing
foreground quasars by spectroscopic means.

In this paper we describe such a search of the field surrounding
Q0302$-$003 using the Focal Reducer/Low Dispersion Spectrograph (FORS)
on the ESO VLT, and the resulting discovery of a nearby quasar closely
coincident in redshift with the isolated \ion{He}{ii} low opacity
feature seen at $z\simeq 3.056$ toward that object.

\section{Observations}

Our approach was to search for quasar candidates at the redshift of
interest through their Ly$\alpha$ emission by imaging the surrounding
field of Q0302$-$003 using VLT/FORS (Seifert et al. 2000) in a
slitless spectroscopic mode. The \ion{He}{ii} low-opacity gap in
Q0302$-$003 is confined to the redshift range $z\simeq 3.05-3.07$
(Heap et al. 2000; Fig.~\ref{blip}). We therefore employed the 150I
(5.52~{\AA}/pixel) low resolution grism crossed with the 486\_37
(4600-5100~{\AA}) intermediate band filter to isolate Ly$\alpha$ over
the range $2.8<z<3.2$  (see B{\"o}hnhardt (2001) for details of the
FORS observing modes). The use of a filter in addition to the
dispersing element had the dual advantage of reducing the risk of
crowding by limiting the length of the individual spectra while
cutting down the (effectively undispersed) sky background in the
slitless exposures.

Our objective was to explore a $13{\arcmin} \times 13{\arcmin}$ wide
field centered on Q0302$-$003 by means of a $2\times 2$ pattern of
overlapping slitless exposures of 1800~s duration. The first
observations were carried out in Service Mode with FORS1 on UT1/Antu
between January 25 and February 20, 2001 (Program 66.A-0307A). The
bulk of these data were taken under less-than-ideal conditions of
bright sky, high airmass (1.4-2.2) and mediocre seeing
(1.1-2.4\arcsec), and therefore only reached down to magnitude
$V\simeq 22$.  Nevertheless, one of the four pointings readily
revealed the presence of a $V\simeq 20.5$ object displaying a broad
emission line at the position $\alpha = 03^{\rm h} 04^{\rm m} 35\fs37
\ \ \delta = - 00\degr 02\arcmin 50\farcs9$ (J2000), $6\farcm5$
northwest of Q0302$-$003 (Fig.~\ref{discovery}).

\begin{figure*}
\centerline{
\resizebox{0.85\hsize}{!}{\includegraphics[angle=-90]{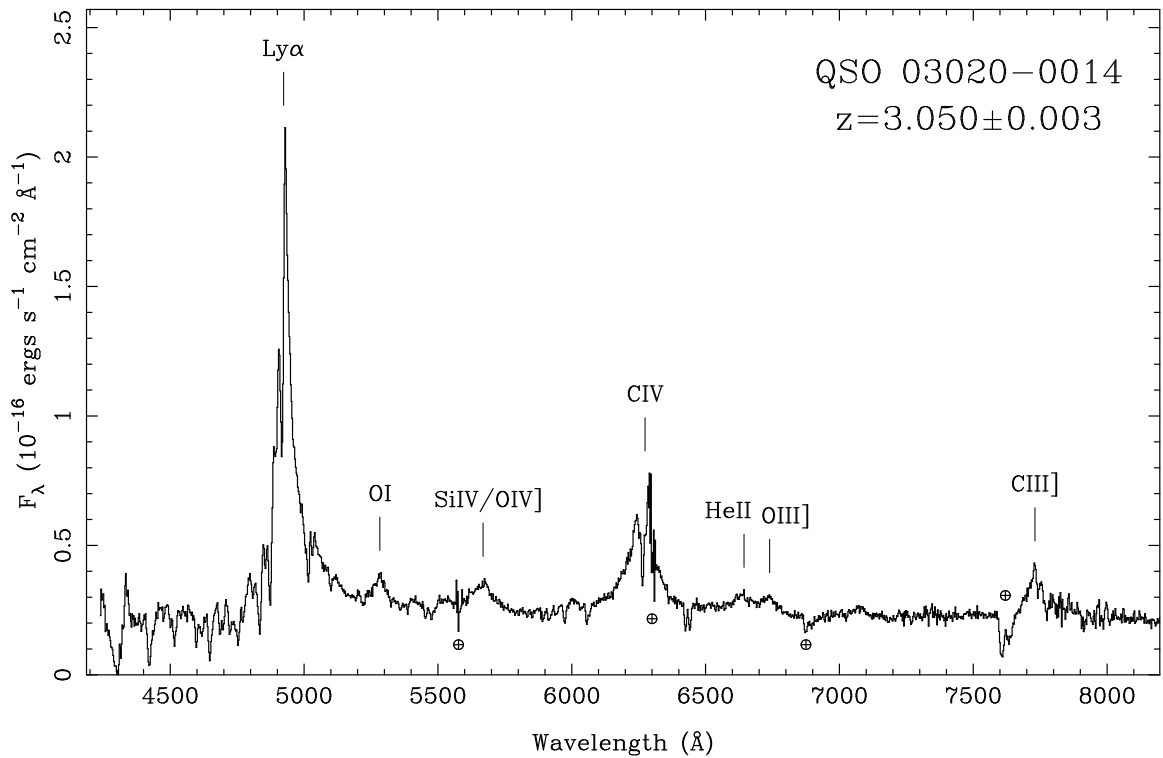}}
}
\caption{VLT/FORS spectrum of QSO\,03020$-$0014 (300V
grism, 3000~s exposure). Emission lines and major atmospheric
artifacts in the spectrum are indicated.}
\label{spectrum}
\end{figure*}

Follow-up spectroscopy of this candidate was obtained with FORS1 on
UT3/Melipal in Visitor Mode on October 15, 2001 (Programme
68.A-0194A). Figure~\ref{spectrum} shows the confirmation spectrum
obtained in 3000~s with the 300V (2.69~{\AA}/pixel) grism and a
$1\arcsec$ wide slit. The spectrum was wavelength calibrated against
the standard He/Ne/Ar/HgCd arc lamps of FORS1 to an accuracy of
$\sigma_\lambda\simeq 0.5$~{\AA}, and photometrically calibrated
against the standard star LTT7987 (Hamuy et al. 1994). The spectrum
confirms that the object is a high redshift quasar, and that the broad
emission line detected in the slitless discovery exposure is indeed
Ly$\alpha$ (Table \ref{lines}). In accord with current IAU conventions
(and to better distinguish it from Q0302$-$003 typographically), we
denote this object QSO\allowbreak\,03020$-$0014 in the
following.\footnote{While preparing this paper, we noted that
QSO\allowbreak\,03020$-$0014 is listed as a point source in the Early
Data Release of the Sloan Digital Sky Survey (Object ID
1970729400402035).}


\begin{table}
\setlength{\tabcolsep}{3pt}
\begin{center}
   \caption[]{Emission lines in QSO\,03020$-$0014}
   \label{lines}
\begin{tabular*}{0.85\linewidth}[h]{ l l l}
\hline\noalign{\smallskip}
\hline\noalign{\smallskip} 
\multicolumn{1}{c}{ID} & 
\multicolumn{1}{c}{Wavelength (\AA)} &
\multicolumn{1}{c}{$z$} \\
\noalign{\smallskip}
\hline\noalign{\smallskip}
Ly$\alpha$& \quad 4925$\pm$15& \quad 3.051$\pm$0.012\\
\ion{O}{i}& \quad 5283$\pm$3& \quad 3.050$\pm$0.002\\
\ion{Si}{iv}/\ion{O}{iv}]& \quad 5671$\pm$15 & \quad 3.052$\pm$0.011\\
\ion{C}{iv}& \quad 6265$\pm$15& \quad 3.044$\pm$0.010\\
\ion{He}{ii}& \quad 6643$\pm$7& \quad 3.050$\pm$0.004\\
\ion{O}{iii}]& \quad 6738$\pm$7& \quad 3.049$\pm$0.004\\
\ion{C}{iii}]& \quad 7728$\pm$10& \quad 3.049$\pm$0.005\\
\noalign{\smallskip}\hline\noalign{\smallskip}
\end{tabular*}
\end{center}
\end{table}


Several intervening absorption line systems can be identified in our
spectrum of QSO\allowbreak\,03020$-$0014. A resolved \ion{Mg}{ii}
doublet matching several \ion{Fe}{ii} lines at shorter wavelength is
clearly seen at $z_{\rm abs}=1.298$. A second weaker \ion{C}{iv}
system is probably present at $z_{\rm abs}=2.910$. Of more interest to
the problem at hand, however, is the strong $z_{\rm abs}\approx z_{\rm
em}$ absorption in Ly$\alpha$, \ion{N}{v} and \ion{C}{iv} evident in
Fig. \ref{spectrum}. This classical high ionization system appears at
$z_{\rm abs}\approx 3.046-3.048$, $\Delta v\simeq
-150-300$~km~s$^{-1}$ shortward of the emission line redshift.

The presence of the $z_{\rm abs}\approx z_{\rm em}$ system (and the
telluric absorption affecting the \ion{C}{iii}] line) hampers the
determination of the precise redshift of QSO\allowbreak\,03020$-$0014.
This is reflected in the large uncertainties listed for several of the
emission lines in Table \ref{lines}. Our adopted redshift of
$z=3.050\pm0.003$ is heavily weighted toward the value indicated by
the well-detected  \ion{O}{i} $\lambda1304$ line. This
choice is also well motivated astrophysically since this low-ionization
line does not suffer from the well-known systematic
blueshifts exhibited by the higher excitation Ly$\alpha$, \ion{C}{iv},
\ion{Si}{iv}/\ion{O}{iv}] and \ion{C}{iii}] lines and therefore provides
the best estimate of the true systemic redshift of
QSO\allowbreak\,03020$-$0014 (Gaskell 1982; Tytler \& Fan 1992;
McIntosh et al. 1999).

With the aim of obtaining an accurate measure of the brightness of
QSO\allowbreak\,03020$-$0014, direct imaging of the object was carried
out in the FORS1 Bessel $B$, $V$ and $R$ filters. These data were
calibrated against the Landolt (1992) standard fields Mark-A and
SA98-670, yielding $B=21.51\pm 0.05$, $V=20.51\pm 0.05$ and
$R=20.20\pm 0.04$, in agreement with our spectrophotometric
calibration.

The initial January/February 2001 Service Mode data also uncovered an
at the time undocumented vignetting issue with FORS in slitless mode
that results in a usable field of view of $5\farcm3 \times 6\farcm8$
for our particular setup, compared to the nominal $6\farcm8 \times
6\farcm8$ assumed. Consequently, a vertical $\simeq 2\arcmin$ wide
strip of sky was missed in the $2 \times 2$ pattern of pointings used
to map the field surrounding Q0302$-$003. For completeness, this gap
was patched through two additional slitless FORS exposures during the
October 2001 Visitor Mode observing run. These observations were
obtained under very good observing conditions (seeing 0.5-0.6\arcsec),
and consequently reach approximately a magnitude deeper than the
initial Service Mode observations. The supplementary slitless
exposures uncovered a second faint broad emission line object in the
field at the position  $\alpha = 03^{\rm h} 04^{\rm m} 45\fs94 \ \
\delta = - 00\degr 11\arcmin 38\farcs2$ (J2000). However, real time
follow-up spectroscopy revealed the object to be a foreground quasar
(QSO\allowbreak\,03022$-$0023) at the `uninteresting' redshift
$z\simeq 2.14$ (\ion{C}{iv} in band).

\section{Discussion}

\subsection{Coincidence in redshift}

\begin{figure}
\centerline{
\resizebox{0.95\hsize}{!}{\includegraphics[angle=-90]{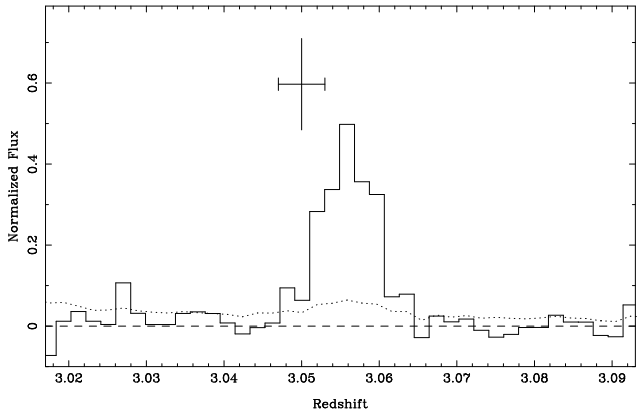}}
}
\caption{Comparison in redshift between the measured emission redshift
of QSO\allowbreak\,03020$-$0014 (cross) and the isolated low opacity
\ion{He}{ii} feature at $z\simeq 3.056$ seen in the STIS spectrum of
Q0302$-$003 (binned full line). The dotted curve denotes the 1$\sigma$
statistical uncertainty per pixel of the STIS spectrum.}
\label{blip}
\end{figure}

In Fig.~\ref{blip}, the location of QSO\allowbreak\,03020$-$0014 in
redshift space is compared to the detailed profile of the \ion{He}{ii}
opacity gap seen at $z\simeq 3.056$ toward Q0302$-$003. The shown
segment of the normalized absorption spectrum is based on our own
reduction of the STIS observations of Heap et al. (2000). Our profile of the
\ion{He}{ii} feature is closely similar to that published by Heap et
al. (2000), but does have slightly sharper edges. Following Heap et
al. (2000) we assumed the continuum to be at $F_\lambda\simeq 2.1
\times 10^{-16}$ erg~s$^{-1}$~cm$^{-2}$~{\AA}$^{-1}$ when normalizing
the spectrum.


It is evident that our measured redshift of $z=3.050\pm0.003$ 
places QSO\allowbreak\,03020$-$0014 about $\Delta z\simeq 0.006$ shortward of
the peak of the opacity gap near its blue edge. Given the low surface
density of quasars at $z\sim 3$, this close coincidence in redshift is
remarkable, and strongly suggests a causal connection between the two.

Using the quasar luminosity function of Pei (1995), we calculate that
the chance probability of a quasar brighter than $V\simeq 20.5$ having
a redshift within $\Delta z \la 0.006$ of the  peak of the
\ion{He}{ii} feature being located within $6\farcm5$ of Q0302$-$003 is
$p\simeq 1 \times 10^{-3}$.

More conservatively, our two slitless spectroscopic observation runs
together covered a $13{\arcmin} \times 13{\arcmin}$ area surrounding
Q0302$-$003; just under half of which was surveyed to a depth of
$V\simeq 23$ and the remainder to $V\simeq 22$. The a priori
probability of our finding a quasar at $3.04 < z < 3.08$ anywhere
within the area surveyed to this depth is $p\simeq 3 \times 10^{-2}$.

As a consistency check, we note that the total number of quasars found
through our slitless observations is in good accord with
expectations. The 4600-5100~{\AA} filter employed in our observations
is capable of detecting high redshift quasars in Ly$\alpha$ over the
range $2.8<z<3.2$; in \ion{C}{iv} over the range $2.0<z<2.3$; and in
\ion{C}{iii}] over the range $1.4<z<1.7$. The predicted number of
quasars of these three categories in our exposures are, respectively,
$\simeq 0.7$, $\simeq 0.5$ and $\simeq 0.2$, which is entirely
consistent with the one Ly$\alpha$ and one \ion{C}{iv} quasar actually
detected.

In other words, the successful result of our search of the field
surrounding Q0302$-$003 needs to be gauged against the $p\simeq 50\%$
a priori probability of our having detected a Ly$\alpha$ quasar
anywhere in our band, compounded by the $p\simeq 9\%$ probability
that the object falls within $\Delta z\simeq\pm 0.02$ of the peak of
the opacity gap of interest.

The above considerations suggest that the null hypothesis that the
close coincidence in redshift between QSO\allowbreak\,03020$-$0014 and
the \ion{He}{ii} absorption gap seen in the spectrum of Q0302$-$003 is
merely a chance occurrence can be rejected at the $\ga 95$\%
confidence level.

\subsection{Astrophysical interpretation}

In their discovery paper, Heap et al. (2000) discuss the reasons for
believing that the \ion{He}{ii} opacity gap seen in the STIS spectrum
of Q0302$-$003 is not merely a statistical fluctuation in the IGM, but
caused by a nearby source of \ion{He}{ii}-ionizing photons. In this
section, we explore the astrophysical implications of presuming that
QSO\allowbreak\,03020$-$0014 is this source.\footnote{Although
galaxies are likely to be feeble sources of $E>54$~eV photons compared
to quasars, we note that the deep survey of Lyman
Break galaxies in the Q0302$-$003 field carried out by Adelberger et
al. (2002) did not reveal any such objects near the $z \simeq 3.06$
redshift of the gap.} Throughout the discussion, we assume a
cosmological model with $H_0=65$~km~s$^{-1}$~Mpc$^{-1}$,
$\Omega_0=0.3$, $\Lambda_0=0.7$, together with a baryonic density of
$\Omega_b h^2 = 0.019$ and a helium abundance by number of $[\mathrm{
He/H}]=0.08$. In this cosmology a redshift difference $\Delta z$
corresponds to a physical distance along the line of sight at redshift
$z$ given by
\begin{equation}
D_\parallel(z) = {c \over H_0} { {\Delta z} \over { (1+z) (\Omega_0
(1+z)^3 + \Lambda_0)^{1 \over 2} } }
\end{equation}
and an angular separation $\theta$ corresponds to a projected distance
\begin{equation}
D_\perp(z)  =  \theta D_L(z) (1+z)^{-2}
\end{equation}
where $D_L$ is the luminosity distance. A convenient approximate
analytical expression for $D_L$ in flat lambda cosmologies is given by
Pen (1999). All distances and physical parameters quoted below refer
to the epoch $z=3.05$.

The angular separation between QSO\allowbreak\,03020$-$0014 and
Q0302$-$003 of $\theta = 6\farcm5$ corresponds to a distance
$D_\perp\simeq 3.2$~Mpc between QSO\allowbreak\,03020$-$0014
($D_{L}\simeq 27.9$~Gpc) and the line of sight to Q0302$-$003. If
ionizing radiation emitted by QSO\allowbreak\,03020$-$0014 is
responsible for creating the \ion{He}{ii} void, it immediately follows
that the quasar must have been on for at least
\begin{equation}
t_Q > D_\perp/c\simeq 1 \times 10^7~\mathrm{yr}
\end{equation}
prior to our observing it at $z\simeq 3.05$. This is consistent with
current estimates of the typical quasar lifetime, which from other
considerations is inferred to be of the order $10^7 \la t_Q \la 
10^8$~yr (Martini \& Weinberg 2001; Haiman \& Hui 2001; Barger et al.
2001 and references therein). The above dimension for the
\ion{He}{iii} region surrounding QSO\allowbreak\,03020$-$0014, and the
corresponding constraint on the quasar lifetime, are similar to those
inferred from the analysis of the `proximity effect' detected in the
\ion{He}{ii} absorption toward Q0302$-$003 and PKS1935$-$692 caused by
the ionizing radiation emitted by the background quasars themselves
(Hogan et al. 1997; Anderson et al. 1999).

\subsection{Photoionization models}

Heap et al. (2000) proposed a simple photoionization model for the
\ion{He}{ii} absorption spectrum of Q0302$-$003 and its opacity gap.
This model was further developed by Smette et al. (2002), who applied
it to the two similar absorption gaps seen in the spectrum of
HE2347$-$4342. The model starts with the conventional picture of an
intergalactic medium in photoionization equilibrium with an ionizing
background that extends to \ion{He}{ii}-ionizing photon energies
$E>54$~eV. The effect of an ionizing source located near the line of
sight is treated as a localized enhancement of the ionizing background
whose intensity decreases as $R^{-2}$ with distance from the object.
By adjusting the assumed spectrum of the ionizing background, the
relative brightness and hardness of the spectrum of the adjacent
quasar, and its distance from the line of sight, the observed width
and contrast of a given \ion{He}{ii} gap can be reproduced.

Smette et al. (2002) also assumed that the initial \ion{He}{ii}
optical depth in the vicinity of an opacity gap is
$\tau^0_\mathrm{He\,II}\simeq 4.5$, which is akin to assuming that the
helium content of the IGM at $z\sim 3$ is already pre-ionized to the
level $\ion{He}{ii}/\ion{He}{iii} \sim \ion{He}{ii}/\ion{He}{} \sim
10^{-2}$. Our examination of the available STIS data has convinced us
that the \ion{He}{ii} observations do not require that this necessarily
be the case, and that the observations are equally consistent with the
significantly larger values of $\tau_\mathrm{He\,II} \approx 10-100$
expected in the delayed \ion{He}{ii} reionization scenario. In
particular, the high opacity \ion{He}{ii} troughs surrounding the
opacity gaps at $z \ga 2.8$ toward Q0302$-$003, HE2347$-$4342, and
PKS1935$-$692 are in our view all consistent with being truly black;
i.e. show no sign of a statistically significant residual flux having
been detected in their dark portions.\footnote{Specifically, our
statistical analysis of the STIS data reveals that the background
signal on the STIS MAMA detectors displays a larger variance than can
be accounted for by photon statistics alone. Consequently, we believe
that the statistical error on the STIS null flux detection limit has
been underestimated in previous work, and that the inferred
\ion{He}{ii} optical depths in the troughs therefore must be
considered to be strict lower limits.} This, of course, does not prove
that the true \ion{He}{ii} optical depth at $z\ga 2.8$ is not, in
fact, lurking at $\tau_\mathrm{He\,II}\simeq 4.5$ just below the STIS
detection threshold as assumed by Smette et al. (2002). However, if
the reduction in absorbing power due to forest structure of the IGM is
taken into account, such a large \ion{He}{ii} optical depth may be
difficult to reconcile with a conventional intergalactic helium
abundance and low baryon density, unless the bulk of the intergalactic
helium is in the form of \ion{He}{ii} at $z\ga 2.8$ (Jakobsen 1998).
Furthermore, the proposition that \ion{He}{ii} was reionized by $z\sim
3$ implicitly requires the existence of a population of faint
energetic $E>54$~eV sources having a sufficiently large space density
to overcome the extremely short mean path for $E>54$~eV photons
implied by the observed strength of the \ion{He}{ii} opacity at
$z\sim3$ (Miralda-Escud{\'e} et al. 2000).

As we shall argue below, our detection of QSO\allowbreak\,03020$-$0014
does not directly address this issue, except in the sense of
demonstrating that there is no need for invoking other
\ion{He}{ii}-ionizing sources beyond luminous quasars. The observed
brightness of QSO\allowbreak\,03020$-$0014 is sufficient for it to
have created its surrounding \ion{He}{iii} zone `from scratch' in an
IGM that had previously only been exposed to sources having a soft
spectrum capable of ionizing \ion{H}{i} and \ion{He}{i} by $z\sim 6$,
but which left the helium content locked in singly-ionized form.

This issue of the uncertain initial conditions aside, another central
underlying assumption implicit in the Smette et al. (2002) model for
the opacity gaps is that the intergalactic medium in the immediate
vicinity of a luminous quasar -- and the \ion{He}{ii}/\ion{He}{iii}
ratio in particular -- is at all times in photoionization equilibrium
with the instantaneous radiation field. That this is not a very
realistic assumption for the problem at hand can be seen by
considering the time scales for the two processes involved in
determining the \ion{He}{ii}/\ion{He}{iii} ionization balance;
photoionization of \ion{He}{ii} and recombination of \ion{He}{iii}.

From Fig.~\ref{spectrum} and the photometry described above, we
estimate the observed continuum flux of QSO\allowbreak\,03020$-$0014
at Ly$\alpha$ to be $F_\lambda\simeq 3\times 10^{-17}$
erg~s$^{-1}$~cm$^{-2}$~{\AA}$^{-1}$. For an assumed extreme
ultraviolet spectrum $F_\nu\propto\nu^{-\alpha}$ with $\alpha\simeq
1.8$ (Telfer et al. 2002), the corresponding rest-frame luminosity of
QSO\allowbreak\,03020$-$0014 at the \ion{He}{ii}-ionization edge is
$L_{\nu\mathrm{ He\,II}}\simeq 3 \times
10^{29}$~erg~s$^{-1}$~Hz$^{-1}$. With this large a flux, the mean time
between photoionizations experienced by \ion{He}{ii} ions situated at
a distance $D_\perp\simeq 3.2$~Mpc from QSO\allowbreak\,03020$-$0014
is very short:
\begin{equation}
t_i= \Gamma_\mathrm{He\,II}^{-1}\simeq ({ {L_{\nu\mathrm{He\,II}}
\sigma_\mathrm{He\,II}} \over {4\pi D_\perp^2 h (\alpha + 3) }} )^{-1}
\simeq 3 \times 10^6~\mathrm{yr} \label{tion}
\end{equation}
where $\sigma_\mathrm{He\,II} = 1.6 \times 10^{-18}$~cm$^2$ is the
\ion{He}{ii} photoionization cross section at threshold. In
comparison, the mean time for \ion{He}{iii} recombination is extremely
long, and comparable to the age of the Universe at $z\simeq 3.05$:
\begin{equation}
t_r\simeq ( \alpha_\mathrm{He\,III} \bar n_\mathrm{e} )^{-1}
\simeq 2 \times 10^9~\mathrm{yr}
\end{equation}
where $\alpha_\mathrm{He\,III} = 1.6 \times 10^{-12}$~cm$^3$~s$^{-1}$
is the \ion{He}{iii} recombination coefficient and $\bar n_\mathrm{e}
\simeq 1 \times 10^{-5}$~cm$^{-3}$ is the mean electron density in the
ionized IGM. Since  $t_r \gg t_i$, and $t_r \ga t_Q$, it follows that
intergalactic \ion{He}{ii} (and \ion{H}{i} and \ion{He}{i} for that
matter) when exposed to a burst of ionizing radiation from a luminous
quasar such as QSO\allowbreak\,03020$-$0014 is unlikely to reach
conditions of ionization equilibrium with photoionization balancing
recombination on the time scale of interest. The IGM at $z\sim 3$ is,
of course, not of uniform density. However, the fact that the
recombination time is even longer in the underdense regions that fill
the bulk of the volume and dominate the \ion{He}{ii} opacity
(Miralda-Escud{\'e} et al. 2000) only strengthens the argument against
the assumption of photoionization equilibrium, except in the extreme
high density regions. In this case, (re)ionization is effectively a
one-way process, and the key parameter determining the extent of the
ionized region surrounding the quasar is not the instantaneous rate of
ionizing photons, but rather the total dose accumulated over time.

The fourth time scale of interest is therefore the time required for
QSO\allowbreak\,03020$-$0014 to emit one $E>54$~eV  photon for every
helium atom contained within a cone reaching out to distance
$D_\perp$:
\begin{equation}
t_\gamma\simeq  {4\pi \over 3} D_\perp^3 { {\bar n}_\mathrm{He} \over {\dot
N}_\mathrm{He\,II}}\simeq5 \times 10^6~\mathrm{yr} \label{photonbudget}
\end{equation}
where ${\dot N}_\mathrm{He\,II}\simeq L_{\nu\mathrm{He\,II}}/h
\alpha\simeq 2 \times 10^{55}$~photons~s$^{-1}$ is the total output of
$E>54$~eV \ion{He}{ii}-ionizing photons and ${\bar n}_\mathrm{He}
\simeq 8 \times 10^{-7}$~cm$^{-3}$ is the average density of
intergalactic helium atoms. Since $t_\gamma < D_\perp/c\simeq 1 \times
10^7$~yr, it follows that QSO\allowbreak\,03020$-$0014 at its observed
brightness at $z\simeq 3.05$ could comfortably have created a
\ion{He}{iii} region reaching out to the sight line to Q0302$-$003 in
the time required -- even if the \ion{He}{ii} is not pre-ionized at
$z\sim 3$.

Note that the above timing and photon budget arguments are equally
applicable to hydrogen. In particular, since the ratio of $E>13.6$~eV
photons to $E > 54$~eV photons emitted by a quasar having an extreme
ultraviolet spectral index $\alpha\simeq 1.8$ is comparable to the
ratio of hydrogen to helium atoms, the burst of \ion{H}{i}-ionizing
radiation from QSO\allowbreak\,03020$-$0014 should have no trouble
accounting for the dearth of Lyman forest lines seen at the position
of the \ion{He}{ii} opacity gaps -- especially since hydrogen was
almost certainly already re-ionized at $z\sim 3$.

In light of the above considerations, a more realistic simple picture
of the interaction of QSO\allowbreak\,03020$-$0014 with the
surrounding IGM  is that its intense burst of ionizing radiation gives
rise to a \ion{He}{iii} region that expands outward for the duration
of the quasar life at a velocity given by
\begin{equation} 
v = {{dR} \over {dt}}
\simeq { {\dot N}_\mathrm{He\,II} \over {4\pi R^2 n^0_\mathrm{He\,II}
} } \quad\quad\quad\quad\quad v\le c \label{propagation}
\end{equation} 
where $n^0_\mathrm{He\,II}$ is the initial \ion{He}{ii} density
encountered in the given direction at the distance $R$ from the
quasar, and it is understood that $dR/dt$ is bounded by the speed of
light (Cen \& Haiman 2000).

Since the intergalactic medium at $z\sim 3$ is not uniform, but
displays density variations spanning more than three orders of
magnitude on the spatial scales of interest, it is clear that the
ionization front from QSO\allowbreak\,03020$-$0014 in reality will
propagate outward in a highly complicated and inhomogeneous manner
that depends sensitively on the initial conditions. At the observed
brightness of QSO\allowbreak\,03020$-$0014, the ionizing radiation
will propagate to distances $R\simeq D_\perp$ at the speed of light in
directions containing only gas of less than average density
$n^0_\mathrm{He\,II} \la 0.7 {\bar n}_\mathrm{He}$. Locations shadowed
by optically thick high density regions, on the other hand, will be
exposed to a lower total ionizing dose since the ionization front will
arrive later due to it needing to first eat its way through the denser
gas according to (\ref{propagation}). Miralda-Escud{\'e} et al. (2000)
describe how, when viewed on larger scales, this process leads to a
gradual reionization of the IGM where the underdense regions are
reionized first and the overdense regions last.

\subsection{Geometry}

As discussed by Smette et al. (2002), there is no reason to expect
QSO\allowbreak\,03020$-$0014 to irradiate its surroundings in an
isotropic manner. According to the unified model for quasars and AGNs,
the central massive black hole is surrounded by a dense optically
thick molecular torus which confines the escaping radiation to a
$\phi\sim 90\degr$ wide ionization cone on each side of the torus
(Barthel 1989). Such beaming can naturally explain why the
\ion{He}{ii} opacity gap appears offset from the redshift of
QSO\allowbreak\,03020$-$0014. 

\begin{figure}
\centerline{
\resizebox{0.99\hsize}{!}{\includegraphics[angle=-90]{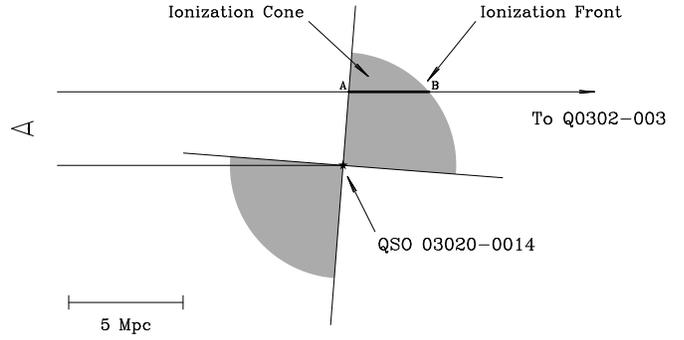}}
}
\caption{Schematic of our proposed model for the geometry of the
\ion{He}{iii} ionization region surrounding
QSO\allowbreak\,03020$-$0014 and its interception by the line of sight
to Q0302$-$003. The extent of the \ion{He}{ii} opacity gap is
indicated by the highlighted line. The scale refers to the epoch
$z=3.05$.}
\label{schematic}
\end{figure}

From Fig.~\ref{blip} we adopt $z_A\simeq
3.051$ as the onset of the opacity gap and $z_B\simeq 3.065$ as its
apparent high redshift edge. For our adopted emission redshift of
$z\simeq 3.050$ the \ion{He}{ii} gap sets in at a distance $D_A\simeq
0.3$~Mpc behind the projected position of QSO\allowbreak\,03020$-$0014
and stops at a distance $D_B\simeq 3.8$~Mpc behind the quasar. This
location of the opacity gap with respect to
QSO\allowbreak\,03020$-$0014 is drawn to scale in
Fig.~\ref{schematic}.

%
%

Also indicated in Fig.~\ref{schematic} is the orientation and extent
of the radiation pattern from QSO\allowbreak\,03020$-$0014 suggested
by the above geometry. It is evident that the free parameters are
tightly constrained. The orientation of the leftmost edge of the top
ionization cone is set by the requirement that the sightline at  $z\la
3.051$ not be illuminated, while the radial extent of the
\ion{He}{iii} zone is fixed by the requirement that the ionization
front has not yet reached beyond the high redshift edge of the gap at
$z \ga 3.065$. Lastly, the requirement that we be able to see and
recognize QSO\allowbreak\,03020$-$0014 as a quasar through the
opposing light cone constrains the opening angle to $\phi\ga 85\degr$
(assuming the top and bottom cones to be symmetrical). Although
admittedly speculative, it is tempting to interpret the strong
$z_\mathrm{abs}\approx z_\mathrm{em}$ absorption detected in our
spectrum of QSO\allowbreak\,03020$-$0014 as evidence that we are
indeed viewing the quasar along a line of sight that grazes the inner
edge of its ionization cone.

At the next level of detail, it is of interest to explore the
approximate shape of the opacity gap anticipated in our proposed
model. In particular, it is worth probing whether it might be possible
to constrain the initial ionization state of the intergalactic
\ion{He}{ii} from the observed profile shape, and thereby address the
outstanding issue of the redshift at which helium reionization
occurred.

It is illustrative to first consider the simplest case of an
\ion{He}{iii} ionization front that propagates at the constant speed
of light over the distances of interest. A key feature of the geometry
shown in Fig.~\ref{schematic} is that different locations along the
line of sight through the opacity gap are not exposed to the same dose
of ionizing radiation at any given time. The ionization front first
reached the line of sight to Q0302$-$003 at position A at the near
side of the opacity gap at a time
$t_A=R_A/c=(D_\perp^2+{D^{2}_A})^{1\over 2}/c\simeq 1.1 \times
10^7$~yr after QSO\allowbreak\,03020$-$0014 first turned on. As the
ionization front propagated further outward, progressively more
distant segments of the sight line become illuminated. The back end of
the gap marks the progression of the ionization front at the time
$t_B=R_B/c=(D_\perp^2+D^{2}_B)^{1\over 2}/c\simeq 1.6\times 10^7$~yr
after turn on. Since the nearer side of the gap is observed at the
later epoch ${\tilde t}_A=t_B+(D_B-D_A)/c\simeq 2.8\times 10^7$~yr, it
follows that at the time of observation the \ion{He}{ii} gas at this
position had been exposed to the ionizing flux from
QSO\allowbreak\,03020$-$0014 for the duration $\Delta
t_E^\mathrm{max}\simeq {\tilde t}_A-t_A\simeq 1.7\times 10^7$~yr. This
value of $\Delta t_E^\mathrm{max}$ becomes larger if the expansion
occurs at sub-luminal speed as per (\ref{propagation}) since more time
is required for the ionization front to propagate from $R_A$ to $R_B$.
Clearly, the typical total exposure time decreases through the opacity
gap from $\Delta t_E =\Delta t_E^\mathrm{max}$ at the near end at
point A to $\Delta t_E =0$ at the far end at point B.

In our proposed geometry, the low redshift edge of the gap reflects
the angular cut-off of the quasar radiation cone shaped by the central
torus, which is unlikely to be perfectly sharp as assumed above, and
could well vary in time. Taking such a gradual extinction of the
ionizing flux into account will make the onset of the gap less sharp
and shift its peak to higher redshift as observed. It is interesting
to note that had the radiation from QSO\allowbreak\,03020$-$0014 not
been hindered in illuminating the front $z \la 3.05$ portion of the
line of sight as in our interpretation, but radiated the line of sight
isotropically, the significant time delay effects would cause the
opacity gap to extend considerably further in front of the quasar than
behind it.

When recombination can be ignored, the residual density of
\ion{He}{ii} ions following an exposure of duration $\Delta t_E$  is
simply
\begin{equation} 
n_\mathrm{He\,II}(\Delta t_E)\simeq n_\mathrm{He\,II}^0
e^{-\Delta t_E / {\bar t}_i} \label{decay}
\end{equation} 
where $n_\mathrm{He\,II}^0$ is the initial \ion{He}{ii} density and
the mean ionization time ${\bar t}_i$ is the reciprocal of the local
ionization rate averaged over $\Delta t_E$
\begin{equation}
{\bar t}_i\simeq {\bar \Gamma}_\mathrm{He\,II}^{-1} = 
({1 \over {\Delta t_E}}\int_{\Delta t_E}
\Gamma_\mathrm{He\,II}\,dt )^{-1} 
\label{avrate}
\end{equation}

Note that since to order of magnitude we have $\Delta t_E \sim
D_\perp/c \sim 10^7$~yr and ${\bar t}_i \sim t_i \sim 10^6$~yr from
(\ref{tion}), we are clearly in a regime where the residual density of
\ion{He}{ii} -- and thereby the observable optical depth
$\tau_\mathrm{He\,II}$ -- is extremely sensitive to even minute
changes in the ratio $\Delta t_E/{\bar t}_i$. Note also that the time
average in (\ref{avrate}) refers to the unknown photon output of
QSO\allowbreak\,03020$-$0014 during a time interval prior to our
observing it at $z\simeq 3.05$.

With these caveats in mind, it is instructive to consider the simplest
case of a constant quasar luminosity and an IGM of uniform density. In
this model ${\bar t}_i$ at a given point on the line of sight is
approximately given by (\ref{tion}) with $R$ substituted for
$D_\perp$, and the observed optical depth corresponding to a given
residual \ion{He}{ii} density is given by the familiar Gunn-Peterson
expression
\begin{equation}
\tau_\mathrm{He\,II} = ({ c \over H_0 } ) { {n_\mathrm{He\,II}(\Delta t_E)
\sigma_l } \over (\Omega_0 (1+z)^3 + \Lambda_0)^{1
\over 2}} \label{GP}
\end{equation}
where $\sigma_l=\lambda_l {{\pi e^2}\over {m_e c^2}} f_{ij} = 1.1
\times 10^{-18}$~cm$^{2}$ is the integrated cross section of the
\ion{He}{ii}~$\lambda304$ transition.

\begin{figure}
\centerline{
\resizebox{0.95\hsize}{!}{\includegraphics[angle=-90]{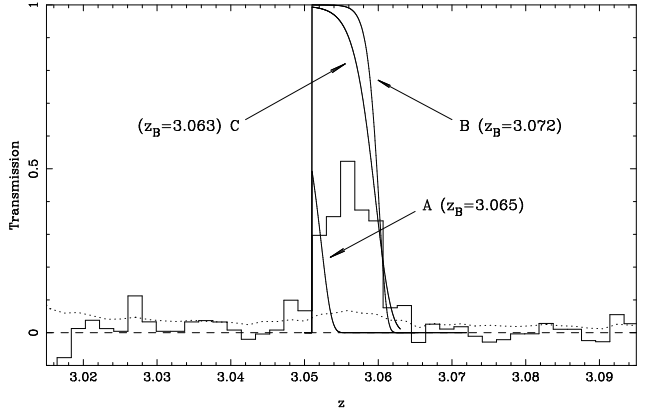}}
}
\caption{Predicted \ion{He}{ii} absorption profiles in the case of a
smooth IGM and a constant quasar output. The labels refer to the
assumed value of $z_B$ and the initial \ion{He}{ii} optical depth (see
text). }
\label{blipfit}
\end{figure}

In Fig.~\ref{blipfit} we plot the predicted gap profiles for three
different combinations of $n_\mathrm{He\,II}^0$ and $R_B$, all
calculated assuming QSO\allowbreak\,03020$-$0014 to have radiated at
its observed brightness at $z\simeq 3.05$ since turn on, and using
(\ref{propagation}) to calculate $R(t)$ and $\Delta t_E$. Case A
assumes the value $z_B\simeq 3.065$ from above and
$n_\mathrm{He\,II}^0 = {\bar n}_\mathrm{He} \simeq 8 \times
10^{-7}$~cm$^{-3}$, corresponding to the fully delayed \ion{He}{ii}
reionization scenario. It is seen that the resulting profile
reproduces the $\sim 50$\% flux decrement of the gap, but not its
width due to the rapid drop-off of the ionizing dose along the line of
sight. However, since in this case the quasar has to create its
\ion{He}{iii} zone in a medium having a huge initial Gunn-Peterson
optical depth of $\tau^0_\mathrm{He\,II} \simeq 2.9 \times 10^3$, the
difference between a gap having $\tau_\mathrm{He\,II} \simeq 1$ and a
completely transparent ($\tau_\mathrm{He\,II} \ll 1$) opacity gap
represents a minute fraction of the total \ion{He}{ii} ionized that
can easily be attributed to residual shielded \ion{He}{ii} in the very
densest regions. The observable $z_B$ is also in practice not
necessarily the redshift at which the ionization front reaches the
most remote part of the line of sight, but rather the redshift at
which the \ion{He}{ii} optical depth of the gap has diminished to a
sufficiently small value for the attenuated quasar flux to rise above
the detection limit. As shown by the second example B in
Fig.~\ref{blipfit}, if one does not attempt to fit the finite depth of
the gap, its width can be reproduced simply by increasing this
parameter to $z_B \simeq 3.072$ -- i.e. by allowing
QSO\allowbreak\,03020$-$0014 to turn on earlier. Example A implicitly
assumes the quasar to have turned on at $z\simeq 3.09$, and having
shone for $t_Q \simeq 4 \times 10^7$~yr prior to our observing it at
$z\simeq 3.05$. In case B the quasar instead turns on at $z\simeq
3.13$, which lengthens the life time to $t_Q \simeq 6 \times 10^7$~yr
and increases the total emitted radiation dose correspondingly. In
other words, even without evoking variations in the brightness of
QSO\allowbreak\,03020$-$0014, the total accumulated ionizing radiation
dose can be matched to any assumed initial value of
$n_\mathrm{He\,II}^0$ simply by adjusting the age of the quasar. This
is illustrated in example C, where $z_B\simeq 3.063$ has been chosen
to match the gap width in the other extreme `pre-ionized' case
$n_\mathrm{He\,II}^0 \simeq 1 \times 10^{-9}$~cm$^{-3}$ corresponding
to $\tau^0_\mathrm{He\,II} \simeq 4.5$ as assumed by Smette et al.
(2002).


Since profiles B and C for our purposes are indistinguishable, this
suggests that more sophisticated modelling will also not be able to
infer the value of $n_\mathrm{He\,II}^0$ from the \ion{He}{ii} gap
profile alone. It is interesting, however, to speculate whether it may
be possible to distinguish between cases B and C by also considering
the \ion{H}{i}-ionizing radiation from QSO\allowbreak\,03020$-$0014.
Since the intergalactic hydrogen is pre-ionized at $z\sim 3$ in both
cases, the additional $E>13.6$~eV flux from
QSO\allowbreak\,03020$-$0014 in the fully delayed \ion{He}{ii}
reionization scenario is expected to reach the line of sight to
Q0302$-$002 faster than the $E>54$~eV flux, in which case the
`transverse' proximity effect on the \ion{H}{i} Lyman forest would be
expected to extend to higher redshift than the \ion{He}{ii} opacity
gap. The Keck spectrum of Q0302$-$002 shown in Fig.~15 of Heap et al.
(2000) does show several weak Ly$\alpha$ lines in the $z\simeq
3.06-3.07$ range, but there is, of course, no way of knowing whether
these systems were high column density systems prior to being
illuminated by QSO\allowbreak\,03020$-$0014 -- just as there is no way
of knowing when QSO\allowbreak\,03020$-$0014 actually turned on and
how bright it was prior to the epoch of observation.

In conclusion, the modelling above, while naive, does serve to
demonstrate that the need to abandon the convenient assumption of
photoionization equilibrium adds several additional unknowns to the
problem of the \ion{He}{ii} gaps, making them exceedingly difficult
to interpret in detail -- beyond the simple photon budget and timing
considerations of Eqs. (\ref{photonbudget}) and
(\ref{propagation}).

\section{Summary and conclusions}

We have identified the adjacent quasar which is almost certainly
responsible for creating the isolated low opacity gap reported by Heap
et al (2000) at $z\simeq 3.056$ in the STIS \ion{He}{ii} absorption
spectrum of Q0302$-$003. The object, QSO\allowbreak\,03020$-$0014,
lies at a distance $\theta = 6\farcm5$ ($D_\perp\simeq 3.2$~Mpc) from
Q0302$-$003, and was uncovered by means of a dedicated search of a
$13{\arcmin} \times 13{\arcmin}$ field centered on Q0302$-$003 using
FORS on the VLT in a filtered slitless spectroscopic mode. Our
follow-up FORS spectroscopy reveals the redshift of
QSO\allowbreak\,03020$-$0014 to be $z \simeq 3.050\pm0.003$, which
places it close to the near edge of the Q0302$-$003 opacity gap in
redshift.

We have presented a plausible model for the shape of the \ion{He}{iii}
ionization zone surrounding QSO\allowbreak\,03020$-$0014. That
QSO\allowbreak\,03020$-$0014 lies slightly in front of the opacity gap in
redshift is interpreted to imply that the \ion{He}{ii}-ionizing flux
from QSO\allowbreak\,03020$-$0014 is confined to two opposing cones as
suggested by unified models of AGNs and quasars. One cone is aimed at
the line of sight to Q0302$-$003 and the object is orientated such
that it can be seen and recognized as a quasar from Earth through its
opposing light cone.

We have also pointed out that the customary assumption of
photoionization equilibrium is very unlikely to be applicable in the
rarefied IGM exposed to the intense and transient radiation field of a
bright quasar such as QSO\allowbreak\,03020$-$0014. In particular,
since the \ion{He}{iii} recombination time is much longer than the
quasar lifetime in all but the most overdense regions of the IGM, the
primary factors determining the extent of the \ion{He}{iii} zone
surrounding a quasar is the initial density of \ion{He}{ii} ions and
the total integrated output of \ion{He}{ii}-ionizing photons emitted
since the quasar first turned on. Both parameters are unknown but at
its observed brightness of $V\simeq 20.5$ at $z\simeq 3.05$,
QSO\allowbreak\,03020$-$0014  could comfortably have ionized a region
large enough to explain the \ion{He}{ii} opacity gap seen toward
Q0302$-$003 in $t_Q \simeq 5 \times 10^7$~yr -- even in the extreme
delayed helium reionization case where the IGM was first reionized by
soft spectrum sources at $z\sim 6$ leaving all the intergalactic
helium locked up in the form of \ion{He}{ii} at $z\simeq 3$.

Our interpretation of the geometry and physics of the
QSO\allowbreak\,03020$-$0014 $+$ Q0302$-$003 system may not bode too
well for the prospect of identifying the neighboring quasars giving
rise to the \ion{He}{ii} opacity gaps detected toward other lines of
sight. Because of the long recombination time scale, some gaps may
well be `fossil' \ion{He}{iii} regions for which the associated quasar
has already turned off. Even if the quasar causing a gap is still
active, we may still not detect and recognize it as such if the light
is confined to an ionization cone orientated perpendicular to the line
of sight. We are presently extending our search technique to the
\ion{He}{ii} gaps seen toward HE2347$-$4342 and PKS1935$-$692, the
outcome of which will reveal how unique QSO\allowbreak\,03020$-$0014
really is.

\begin{acknowledgements}
The staff of the Paranal Observatory are thanked for their helpful and
professional assistance in carrying out the VLT observations described
in this paper.
\end{acknowledgements}


\begin{thebibliography}{}

\bibitem[2002]{adelberger} Adelberger, K. L., Steidel, C. C., Shapely, A. E., \& Pettini, M., 2002 ApJ (in press, astro-ph/0210314)

\bibitem[1999]{anderson_1935} Anderson, S. F., Hogan, C. J., Williams, B. F., \& Carswell, R. F., 1999  AJ, 117, 56

\bibitem[2001]{barger} Barger, A. J., Cowie, L. L., Bautz, et al., 2001, AJ 122, 2177

\bibitem[1989]{bartel} Barthel, P. D., 1989, ApJ 336, 606.

\bibitem[2001]{becker} Becker, R. H., Fan, X., White, R. L., et al., 2001, AJ 122, 2850

\bibitem[2001]{boenhardt} B{\"o}hnhardt, H. (ed.) 2000, FORS1+2 User Manual, VLT-MAN-ESO-13100-1543/2.2

\bibitem[2000]{cenhaiman} Cen, R., \& Haiman, Z., 2000, ApJ 542, L75

\bibitem[1996]{davidsen_1700} Davidsen, A. F., Kriss, G. A., \& Zheng, W., 1996, Nature 380, 47

\bibitem[2001]{georgie} Djorgovski, S. G., Castro, S. M., Stern, D., \& Mahabal, A. A., 2001,  ApJ 560, L5

\bibitem[1982]{gaskell} Gaskell, C. M., 1982, ApJ 263, 79

\bibitem[2001]{haiman} Haiman, Z., \& Hui, L., 2001, ApJ 547, 27

\bibitem[1994]{hamuy} Hamuy, M., Suntzeff, N. B., Heathcote, S. R., et al., 1994, PASP 106, 566 

\bibitem[2000]{heap_0302} Heap, S. R., Williger, G. M., Smette, A., et al. 2000, ApJ 534, 69

\bibitem[1997]{hogan_0302} Hogan, C. J., Anderson, S. F., \& Rugers, M. H., 1997, AJ 113, 1495

\bibitem[1998]{jakobsen_omega} Jakobsen, P., 1998, A\&A 331, 61

\bibitem[1994]{jakobsen_0302} Jakobsen, P., Boksenberg, A., Deharveng, J. M., et al., 1994, Nature 370, 35

\bibitem[2001]{kriss_2347} Kriss, G. A.,  Shull, J. M., Oegerle, W., et al. 2001, Science 293, 5532

\bibitem[1992]{landolt} Landolt, A. U., 1992, AJ 104, 340

\bibitem[2001]{martini} Martini, P., \& Weinberg, D. H., 2001 ApJ 547, 12

\bibitem[1999]{mcintosh} McIntosh, D. H.,  Rix, H.-W., Rieke, M., \& Foltz, C. B., 1999 ApJ 517, L73

\bibitem[2000]{miralda_reion} Miralda-Escud{\'e}, J., Haehnelt, M., \& Rees, M. J., 2000 ApJ 530, 1

\bibitem[1995]{pei} Pei,ÊY. C., 1995, ApJ 438, 623

\bibitem[200?]{pen} Pen, U., 1999, ApJS 120, 49

\bibitem[2002]{penttricci} Pentericci, L., Fan, X., Rix, H. W., et al., 2002 ApJ 123, 2151

\bibitem[1997]{reim_2347} Reimers, D., K{\"o}hler, S., Wisotzki, L., et al., 1997, A\&A 327, 890

\bibitem[2000]{seifert} Seifert, W., Appenzeller, I., F{\"u}rtig, W., et al. 2000, Proc. SPIE 4008, ed. M. Iye, \& A. F. Moorwood, 96

\bibitem[2002]{smette_2347} Smette, A., Heap, S. R., Williger, G. M., et al., 2002,  ApJ 654, 542

\bibitem[2002]{telfer} Telfer, R.C., Zheng, W., Kriss, G. A., \& Davidsen, A. F, 2002, ApJ 565. 773

\bibitem[1992]{tytler&fan} Tytler, D., \& Fan, X. M., 1992, ApJS 79, 1

\bibitem[1995]{tytler_1935} Tytler, D., Fan, X. M., Burles, S., et al., 1995, In:  G. Meylan (ed.) QSO Absorption Lines, Springer, Berlin, p. 289

\end{thebibliography}
\end{document}